# On the properties of solar energetic particle events associated with metric type II radio bursts

Pertti Mäkelä[1,2], Nat Gopalswamy[2], Hong Xie[1,2], Sachiko Akiyama[1,2], Seiji Yashiro[1,2], Neeharika Thakur[1,2]

[1] The Catholic University of America, Washington, District of Columbia, USA

[2] NASA Goddard Space Flight Center, Greenbelt, Maryland, USA

Email: pertti.makela@nasa.gov

**Abstract** Metric type II solar radio bursts and solar energetic particles (SEPs) are both associated with shock fronts driven by coronal mass ejections (CMEs) in the solar corona. Recent studies of ground level enhancements (GLEs), regular large solar energetic particle (SEP) events and filament eruption (FE) associated large SEP events have shown that SEP events are organized by spectral index of proton fluence spectra and by the average starting frequencies of the associated type II radio bursts. Both these results indicate a hierarchical relationship between CME kinematics and SEP event properties. In this study, we expand the investigations to fluence spectra and the longitudinal extent of metric type II associated SEP events including low-intensity SEP events. We utilize SEP measurements of particle instruments on the Solar and Heliospheric Observatory (SOHO) and Solar Terrestrial Relations Observatory (STEREO) spacecraft together with radio bursts observations by ground-based radio observatories during solar cycle 24. Our results show that low-intensity SEP events follow the hierarchy of spectral index or the hierarchy of the starting frequency of type II radio bursts. We also find indications of a trend between the onset frequency of metric type II bursts and the estimated longitudinal extent of the SEP events although the scatter of data points is quite large. These two results strongly support the idea of SEP acceleration by shocks. Stronger shocks develop closer to the Sun.

**Keywords:** coronal mass ejections, type II radio bursts, shocks, solar energetic particle events, space weather

## Introduction

Type II solar radio bursts and large solar energetic particle (SEP) events are both associated with shock fronts driven by coronal mass ejections (CMEs) in the solar corona and interplanetary (IP) space. Type II radio bursts appear in radio dynamic spectra as features whose frequency slowly drifts towards lower frequencies due to decreasing electron density as the shock propagates outwards from the Sun (e.g., Wild and McCready, 1950; Nelson and Melrose, 1985). Type II radio bursts are associated with CME-driven shocks that accelerate electrons generating Langmuir waves, which are then converted to radio waves detected as type II radio emissions at the fundamental (F) and/or second harmonic (H) of the plasma frequency in the shock upstream region. Ground-based observations of type II solar radio bursts and observations of proton events in the interplanetary (IP) space during January 1966–June 1968 established a correlation between the metric type II radio bursts and high-energy proton events (Švetska and Fritzová-Švestková, 1974). Later, Cane and Stone (1984) found a similar association between solar energetic particle (SEP) events and longer-wavelength IP type II radio bursts that can be detected by radio receivers on spacecraft. With the advent of the Radio and Plasma Wave Experiment (WAVES, Bougeret et al., 1995) on board the Wind spacecraft, type II bursts are observed over the entire inner heliosphere and the degree of their SEP association depends on the wavelength range of the bursts (Gopalswamy et al. 2008). Statistical studies show that ~80% of 20 MeV SEP events are associated with metric type II bursts (Cliver et al., 2004). Papaioannou et al. (2016) studied 174 SEP events observed during 1997–2013 and found that 129 out of 174 (74%) SEP events were associated with type II bursts, while additional 10 SEP events showed indications of type II bursts. In the impulsive SEP events particles are accelerated in solar flares that do not have CME-driven shocks producing type II bursts. Type II emission also requires that the CME-driven shock accelerates electrons, whereas SEP events require acceleration of ions. The efficiency of electron and ion acceleration by a shock could differ, which will affect the SEP event association with type II radio bursts. The intensity of type II emission must also be above the background level of all the other concurrent radio emissions at the same wavelength range, before the radio receiver can detect it.

The white-light observations of CMEs near the Sun by the Skylab mission from May 1973 to February 1974 provided the first confirmation that fast CMEs were associated with SEP events (Kahler et al., 1978). Current understanding is that energetic particles observed during large SEP events are accelerated at CME-driven shocks near the Sun (see e.g., Reames, 1999; Desai and Giacalone, 2016). Following the SEP event definition by the National Oceanic and Atmospheric Administration (NOAA), large (also called major) SEP events are defined as particle events where the peak flux in the >10 MeV integral proton channel of the Geostationary Operational Environmental Satellite (GOES) exceeds 10 pfu (1 pfu = 1 particle flux unit = $1/cm^2/sec/sr$).

Most fast CMEs producing large SEP events are launched from the solar corona above active regions (ARs) concurrently with an intense flare eruption occurring below the associated CME, indicating that reconnection of strong magnetic fields within ARs is the common source of energy for both flares and CMEs. As an exception to this general rule, Kahler et al. (1986) reported on the 1985 December 5 filament eruption (FE) without an intense concurrent flare that occurred outside any AR but was associated with a large SEP event observed by a particle instrument on the International Sun-Earth Explorer (ISEE) 3 spacecraft. The SEP event had an unusually soft energy spectrum (power-law index ≈ 4) compared to the energy spectra of other large SEP events. In addition, no metric type burst was reported, only a longer-wavelength IP type II burst was detected suggesting shock formation at a larger distance from the Sun. Gopalswamy et al. (2015) found four new FEs outside ARs that were associated with large SEP events and IP type II radio bursts. Only one of the FEs was associated with a metric type II burst, and the spectral indexes of the SEP events in the 10–100 MeV energy range were larger than 4. Expanding on these results, Gopalswamy et al. (2016) used GOES and the Solar, Anomalous, and Magnetospheric Particle Explorer (SAMPEX; Baker et al., 1993) observations to study the fluence spectra of ground level enhancement (GLE) events, well-connected large SEP events and FE-associated SEP events during cycles 23 and 24. GLE events, large SEP events, and FE-associated SEP events are organized by spectral index of proton fluence spectra and by the average starting frequencies of the associated type II radio bursts. These results indicate a hierarchical relationship between CME kinematics and SEP event properties.

Another extensively studied topic of SEP research is the longitudinal extent of SEP events. A few early observations of SEPs had already indicated that particles from eruptions far behind the solar western limb can be detected near Earth (Dodson et al., 1969; Dodson and Hedeman, 1969). Other more recent studies reached similar conclusions (e.g., Cane, 1996; Torsti et al., 1999; Cliver et al., 2005). After the launch of the Solar Terrestrial Relations Observatory (STEREO; Kaiser et al., 2008) mission in 2006, the multi-spacecraft observations of SEP events showed indisputably that solar eruptions occurring ~180° away from the observer's magnetic connection point at the Sun can still produce detectable SEP events, thus proving the circumsolar extent of those SEP events (e.g., Dresing et al., 2012; Lario et al., 2014, 2016; Richardson et al., 2014; Gómez-Herrero et al., 2015). The longitudinal extent of SEP events is variable from event to event. It has been known for a long time that the speed of a CME and the peak intensity of the associated SEP event are correlated (e.g., Kahler, 2001, and references therein). The width of a CME is also known to correlate with SEP intensity (Kahler et al., 1984).

Gopalswamy et al. (2008) showed type II bursts at decameter-hectometric (DH) wavelengths are the best indicators for large SEP events among all type II radio bursts occurring at various wavelength ranges, but the limitation of DH type II radio bursts is that they give only a short or no lead time for prediction because SEPs arrive at 1 AU around the same time as the DH type II bursts are observed. It is also known that SEP-producing CMEs are wider than non-SEP-producing CMEs (Gopalswamy et al., 2008; Papaioannou et al., 2016). The type II burst and SEP event association rates also increase with CME speed and width (Gopalswamy et al., 2008).

We have expanded our previous research of hierarchical relations of spectral index and type II starting frequency in large SEP event to include small SEP events observed by the Energetic and Relativistic Nuclei and Electron (ERNE; Torsti et al., 1995) experiment on the Solar and Heliospheric Observatory (SOHO; Domingo et al., 1995) spacecraft. We have studied the longitudinal extent of multi-spacecraft SEP events observed by particle instrument on the SOHO and STEREO spacecraft reported by Richardson et al. (2014) in association with metric type II radio bursts. The main goal of the study is to check if type II radio bursts can be used to predict the energy spectrum and longitudinal extent of SEP events in the near-Earth space.

# Data analysis

For our analysis of type II radio bursts in solar cycle 24 we have used radio measurements provided by ground-based radio observatories, including those of the Radio Solar Telescope Network (RSTN; https://www.ngdc.noaa.gov/stp/space-weather/solar-data/solar-features/solar-radio/rstn-spectral/), the Compound Astronomical Low cost Low frequency Instrument for Spectroscopy and Transportable Observatory network (CALLISTO; https://www.e-callisto.org/), Culgoora Solar Radio Spectrograph (https://www.sws.bom.gov.au/Solar/2/2/), Green Bank Solar Radio Burst Spectrometer (GBSRBS; https://www.astro.umd.edu/~white/gb/), Izmiran Radio Spectrograph (https://www.izmiran.ru/stp/lars/) and Hiraiso Radio Spectrograph (HiRAS; https://sunbase.nict.go.jp/solar/denpa/). The dynamic spectra in the metric wavelength range were visually investigated to estimate the onset frequencies of the type II radio bursts. The dynamic spectra were also used to examine if the measured onset frequency corresponded to fundamental or harmonic emission in events where the type II burst had a clear fundamental-harmonic (F-H) structure.

For the analysis of small SEP events, defined as SEP events where the peak flux of >10 MeV particles measured by the Geostationary Operational Environmental Satellite program (GOES) satellites remains below 10 pfu during the event, we used particle measurements provided by the High-Energy Detector (HED) of SOHO/ERNE instrument. The ERNE/HED measures protons in the energy range of 13.8 MeV-120 MeV. ERNE/HED can detect much smaller increases of proton flux than the GOES particle instruments, which suffer from elevated background level of particle flux. Because the flux of the SEPs depends on the magnetic connection between the particle source and the observing spacecraft, we included in our analysis only those small SEP events that had the source location of the associated solar eruptions in the western hemisphere or just behind the western limb. The source locations were estimated from EUV images taken by the Atmospheric Imaging Assembly (AIA; Lemen et al., 2012) on the Solar Dynamic Observatory (SDO; Pesnell et al., 2012) and the Extreme UltraViolet Imager (EUVI; Wülser et al., 2004) of the Sun Earth Connection Coronal and Heliospheric Investigation (SECCHI; Howard et al., 2008) on the STEREO spacecraft.

After inspecting solar radio emission and high-energy proton measurements, we found a total of 17 well-connected small SEP events with a metric type II radio burst. The selected small SEP events are listed in Table 1, where the first three columns give the date, time and starting frequency in units of MHz of the metric type II radio burst. The fourth column indicates if the listed frequency is fundamental (F) or uncertain (U) emission, i.e., if the type II radio burst had a F-H structure or not. The columns 5-8 give the start and end dates and times of the associated small SEP event. The column 9 lists the spectral index ($\gamma$) of the SEP fluence spectrum and the column 10 gives how many measurement points were used in fitting the power law. The columns 11-15 list the date, start time and peak time of the associated flare together with the significance class of GOES soft X-ray emission and the location of the eruption in heliographic coordinates. The columns 16-22 list the first observation date and time of the associated CME observed by the Large Angle and Spectrometric Coronagraph (LASCO; Brueckner et al., 1995) on SOHO together with the initial acceleration ($a_{ini}=V_{sp}/(t_{peak}-t_{start})$) and ($V_{ini}=(h_2-h_1)/(t_2-t_1)*V_{sp}/V_{sky}$), sky-plane ($V_{sky}$) and space ($V_{sp}$) speed of the CME. Times $t_{start}$ and $t_{peak}$ are the flare onset and peak times. Times $t_1$ and $t_2$ are the times of the first two CME height measurements $h_1$ and $h_2$ in the LASCO field of view (FOV), correspondingly. The space speed $V_{sp}$ is a projection corrected speed for non-halo CMEs and for halo CMEs it is taken from the SOHO/LASCO Halo CME catalog. Gopalswamy et al. (2012) discuss different methods for estimating CME initial acceleration and speed.

For the analysis of the longitudinal extent of the SEP events we used the list of >25 MeV proton events published by Richardson et al. (2014), which includes particle events observed from the launch of the STEREO mission in 2006 until the end of 2013. They analyzed proton observations provided by the High Energy Telescopes (HETs; von Rosenvinge et al., 2008) on the twin STEREO spacecraft together with observations from the Electron Proton and Helium Instrument (EPHIN; Müller-Mellin et al., 1995) and the ERNE both on the SOHO spacecraft near Earth. They could identify a total of 209 individual proton events that were detected by either one or multiple spacecraft. From this list we selected SEP events associated with a metric type II burst. We excluded SEP events where at least one spacecraft had an elevated background. Because type II emission at metric wavelengths originates from CME-driven shocks propagating in the low solar corona and can be measured only by ground-based observatories, the metric type II bursts can be detected if the solar source of the eruption is located on the solar disk or sufficiently close behind the solar limb. Therefore, we included only 99 events, where the longitude of the solar source (flare) was in the range E95°-W95°. The source locations are mostly on the northern hemisphere, because the study period covers the beginning of solar cycle 24 where solar activity in the northern hemisphere increased more quickly and peaked earlier than in the southern hemisphere.

Our analysis found that 27 out of the 28 (94%) 3-spacecraft (3 S/C) SEP events were associated with a metric type II burst. Correspondingly, 21 out of the 36 (58%) 2-spacecraft (2 S/C) events and 19 out of the 35 (54%) 1-spacecraft (1 S/C) events were associated with a metric type II burst. On average, 67 events of the 99 (68%) selected SEP events had a metric type II burst. We also checked for the existence of a DH and kilometric type II radio burst in the 3, 2 and 1 S/C SEP events with type II burst, and found that 24 out of 27 (89%), 17 out of 21 (81%) and 14 out of 19 (74%) had also a DH/km type II burst, respectively. The corresponding numbers for the 32 SEP events without a metric but with a DH/km type II burst were 1 out of 1 (100%), 7 out of 15 (47%) and 6 out of 16 (38%) events. The source locations of the selected SEP events with (red and green circles) and without (blue and magenta circles) a metric type II radio burst are marked in Fig. 1. The green color indicates that the SEP event with a metric type II burst is also associated with a DH or km type II burst, whereas the magenta color indicates SEP events without a metric type II burst that are associated with a DH or km type II burst. The size of the circle indicates the number of observing spacecraft. A total 63 (63%) of the 99 SEP events in the longitude range E95°-W95° were associated with a DH/km type II burst. One should note that the set of SEP events includes other than large SEP events, therefore the DH/km type II association rate is lower than that for large SEP events.

**Table 1** Parameters of radio bursts, small SEP events, flares and CMEs.

| Metric Type II Burst | | | | SEP Event | | | | | | Flare | | | | | CME | | | | | | |
|---|---|---|---|---|---|---|---|---|---|---|---|---|---|---|---|---|---|---|---|---|---|
| Date | Time UT | Freq MHz | | Date | Start UT | Date | End UT | $\gamma$ | # | Date | Start UT | Peak UT | Class | Location ° | Date | Time UT | W deg | $a_{ini}$ km/s$^2$ | $V_{ini}$ km/s | $V_{sky}$ km/s | $V_{Sp}$ km/s |
| (1) | (2) | (3) | (4) | (5) | (6) | (7) | (8) | (9) | (10) | (11) | (12) | (13) | (14) | (15) | (16) | (17) | (18) | (19) | (20) | (21) | (22) |
| 2009/12/22 | 04:57 | 85 | F | 12/22 | 05:30 | 12/24 | 19:10 | 2.56 | 6 | 12/22 | 04:50 | 04:56 | C7.2 | S26W46 | 12/22 | 05:54 | 47 | 1.14 | 450 | 318 | 411[a] |
| 2010/06/12 | 00:57 | 105 | F | 06/12 | 01:45 | 06/16 | 22:40 | 3.14 | 7 | 06/12 | 00:30 | 00:57 | M2.0 | N23W43 | 06/12 | 01:31 | 119 | 0.41 | 646 | 486 | 660[a] |
| 2010/08/18 | 05:51 | 60 | U | 08/18 | 07:00 | 08/22 | 01:30 | 4.46 | 7 | 08/18 | 04:45 | 05:48 | C4.5 | N17W101 | 08/18 | 05:48 | 184 | 0.40 | 1493 | 1471 | 1498[a] |
| 2011/05/11 | 02:27 | 75 | F | 05/11 | 02:10 | 05/14 | 00:00 | 3.27 | 7 | 05/11 | 02:23 | 02:43 | B8.1 | N19W51 | 05/11 | 02:48 | 225 | 0.76 | 794 | 745 | 917[a] |
| 2011/08/02 | 06:40 | 45 | F | 08/02 | 06:40 | 08/03 | 15:08 | 2.79 | 7 | 08/02 | 05:58 | 06:19 | M1.4 | N14W15 | 08/02 | 06:36 | 268 | 2.22 | 2800 | 712 | 2431[a] |
| 2013/04/28 | 20:18 | 41 | F | 04/28 | 22:20 | 05/01 | 01:20 | 3.03 | 5 | 04/28 | 20:10 | 20:17 | C4.4 | S18W37 | 04/28 | 20:48 | 91 | 1.88 | 630 | 497 | 788[a] |
| 2013/05/02 | 05:06 | 80 | F | 05/02 | 09:10 | 05/05 | 05:20 | 3.20 | 6 | 05/02 | 04:58 | 05:10 | M1.1 | N10W26 | 05/02 | 05:24 | 99 | 1.90 | 1460 | 671 | 1370[a] |
| 2013/08/17 | 18:56 | 63 | U | 08/17 | 20:45 | 08/19 | 23:50 | 4.74 | 6 | 08/17 | 18:49 | 19:33 | M1.4 | S05W30 | 08/17 | 19:12 | 360 | 0.54 | 822 | 1202 | 1418[b] |
| 2013/12/07 | 07:24 | 120 | F | 12/07 | 11:45 | 12/10 | 02:45 | 3.62 | 6 | 12/07 | 07:17 | 07:29 | M1.2 | S16W49 | 12/07 | 07:36 | 360 | 1.62 | 1390 | 1085 | 1165[b] |
| 2014/06/12 | 22:00 | 57 | U | 06/12 | 23:40 | 06/17 | 13:15 | 3.18 | 7 | 06/12 | 21:34 | 22:16 | M3.1 | S20W55 | 06/12 | 22:12 | 186 | 0.32 | 698 | 684 | 810a |
| 2014/08/25 | 15:08 | 45 | F | 08/25 | 16:50 | 08/28 | 20:45 | 3.45 | 7 | 08/25 | 14:46 | 15:11 | M2.0 | N05W36 | 08/25 | 15:36 | 360 | 0.46 | 932 | 555 | 697[b] |
| 2015/09/20 | 18:16 | 40 | U | 09/20 | 18:35 | 09/24 | 07:50 | 3.60 | 7 | 09/20 | 17:32 | 18:03 | M2.1 | S20W24 | 09/20 | 18:12 | 360 | 0.78 | 1171 | 1239 | 1458[b] |
| 2015/11/04 | 13:43 | 85 | F | 11/04 | 15:30 | 11/06 | 05:00 | 2.86 | 7 | 11/04 | 13:31 | 13:52 | M3.7 | N09W04 | 11/04 | 14:48 | 360 | 0.78 | 988 | 578 | 987[b] |
| 2016/02/11 | 20:35 | 70 | F | 02/12 | 01:15 | 02/15 | 17:40 | 3.74 | 5 | 02/11 | 20:18 | 21:03 | C8.9 | N11W07 | 02/11 | 21:17 | 360 | 0.43 | 1827 | 719 | 1174[b] |
| 2016/03/16 | 06:45 | 85 | U | 03/16 | 07:35 | 03/21 | 13:00 | 2.99 | 7 | 03/16 | 06:34 | 06:46 | C2.2 | N12W88 | 03/16 | 07:00 | 154 | 0.82 | 1043 | 592 | 592[a] |
| 2016/04/18 | 00:30 | 80 | U | 04/18 | 02:05 | 04/21 | 13:20 | 4.12 | 6 | 04/18 | 00:14 | 00:29 | M6.7 | N12W62 | 04/18 | 00:48 | 162 | 1.35 | 1474 | 1084 | 1213[a] |
| 2017/04/03 | 14:25 | 80 | F | 04/03 | 19:10 | /04/05 | 04:00 | 3.71 | 4 | 04/03 | 14:19 | 14:29 | M5.8 | N16W79 | /04/03 | 14:48 | 80 | 0.80 | 764 | 471 | 479[a] |

[a] $V_{sp}=V_{sky}/\cos(angle\_from\_sky\_plane)$; [b] SOHO/LASCO Halo CME catalog https://cdaw.gsfc.nasa.gov/CME_list/halo/halo.html

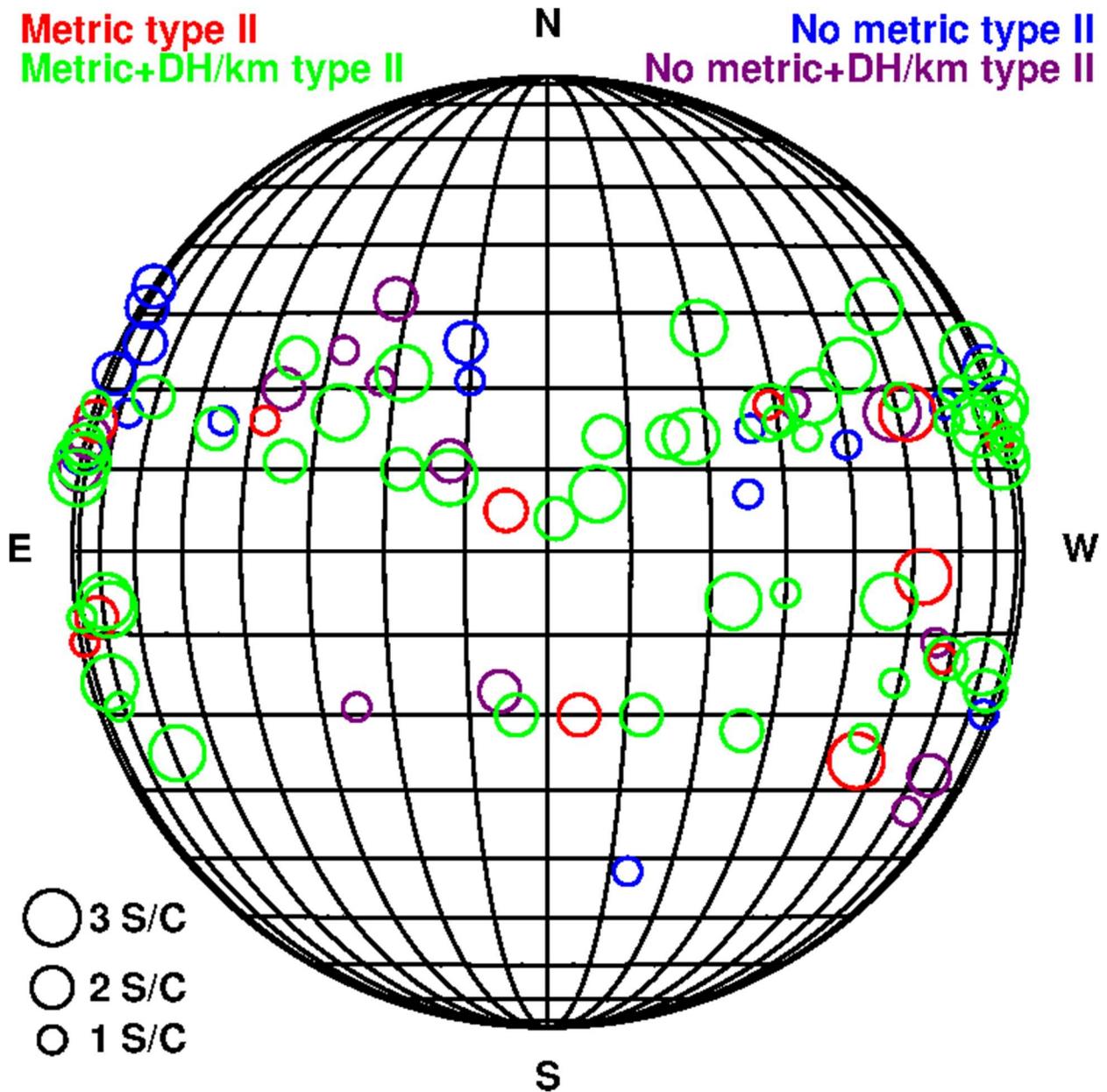

**Figure 1** Solar source locations of SEP events with (red and green circles) and without (blue and magenta circles) a metric type II radio burst. The green color indicates that the SEP event with a metric type II burst is also associated with a DH or km type II burst, whereas the magenta color

indicates SEP events without a metric type II burst that are associated with a DH or km type II burst. The size of the circle indicates the number of observing spacecraft (S/C).

In order to estimate the extent of the SEP events we have used a simple geometrical method based on the longitude of the observing spacecraft. Because charged particles follow magnetic field lines, we have shifted spacecraft longitude according to the Parker spiral for the average 400 km/s solar wind in order to estimate the angular distance of the observing spacecraft from the solar source location (flare) from where the SEP source expands. This method assumes that particles are released on the magnetic field line connecting back to the observing spacecraft but does not specify how particles reached this field line, i.e., due to cross-field diffusion or due to expanding shock front. Using this information, the estimated upper limit for the longitudinal extent of the SEP event for 1-spacecraft events is given by the longitudinal separation angle between the nearest no-SEP-observing spacecraft and the source (flare). Similarly, the estimated upper limit for the longitudinal extent of the 2-spacecraft SEP event is given by the longitudinal separation angle between the third no-SEP-observing spacecraft and the flare. In the case of 3-spacecraft events, we can estimate only the lower limit of the longitudinal extent, which is given by the longitudinal separation angle between the farthermost SEP-observing spacecraft and the source. All estimates of SEP propagation in interplanetary space suffer from the fact that we cannot know the actual configuration of the Parker field, especially when there are intervening disturbances propagating between the particle source and the observer. Because of this inaccuracy and because from the multi-spacecraft observations we can estimate only the upper or lower limits for the longitudinal extent, we do not consider the changes in field line configuration due to differences in solar wind speed between each event and spacecraft location, which we have deemed to be second-order corrections for our analysis. If we assume that the longitudinal extent of SEP events is symmetrical relative to the solar source (flare), then the estimated separation angle is half of the total longitudinal width of the SEP event.

## Results and discussion

Figure 2 shows the distribution of power-law index, γ, fitted to the fluence spectra of the selected 17 small SEP events observed by ERNE. The mean value of the spectral index is 3.42 and the median value is 3.27. Gopalswamy et al. (2016) studied the fluence spectra of GLEs, regular large SEP events from the western hemisphere and FE-associated events in solar cycles 23 and 24. They found that the mean (median) spectral indexes for the cycle 23 and 24 events combined were 2.68 (2.67) for 18 GLEs, 3.83 (3.77) for 60 regular large SEP events and 4.89 (4.92) for 8 FE-associated events. Our results for 17 small SEP events are slightly higher but close to those of the regular SEP events in Gopalswamy et al. (2016). Our results show that the fluence spectral index of the small SEP events fit excellently to the 3-group hierarchical structure suggested by Gopalswamy et al. (2016), i.e., large SEP events associated with FEs outside active regions have high fluence spectral indexes, regular large SEP events have intermediate index values, and GLE events have low index values. Small SEP events fall between regular and FE-associated large SEP events.

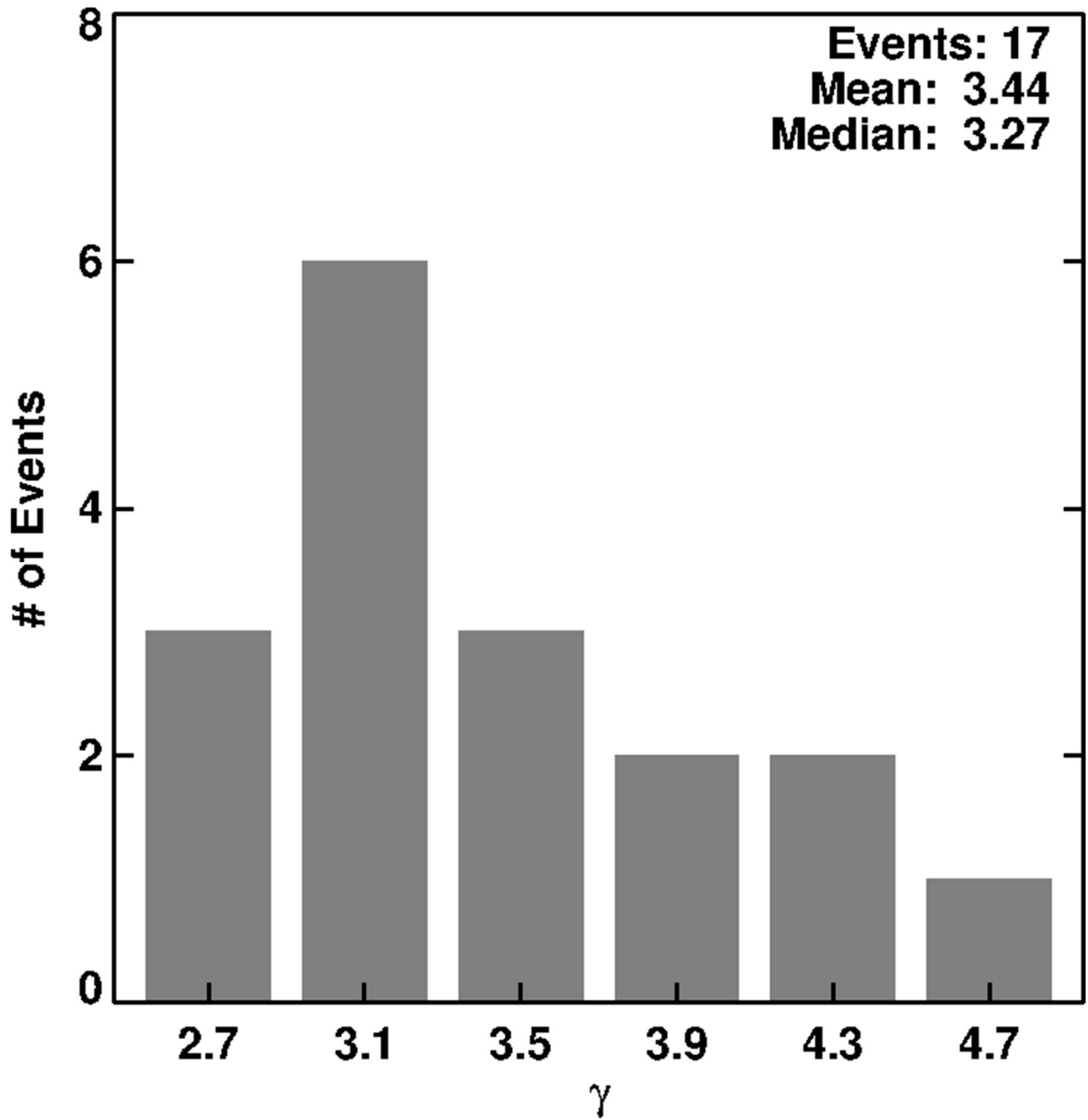

**Figure 2** The distribution of the fluence spectral index, γ, for small SEP events.

Gopalswamy et al. (2016) found also a clear anticorrelation between fluence spectral index and CME space speed: the average spectral index and CME space speed were for FE-associated SEP events (γ=4.89; $V_{sp}$=1195 km/s), regular SEP events (γ=3.86; $V_{sp}$=1787 km/s) and GLEs (γ=2.61; $V_{sp}$=2295 km/s), respectively. The average space speed of CMEs causing small SEP events, V=1063 km/s, is clearly below the lowest average CME speed of the FE-associated CMEs. However, this discrepancy with the anticorrelation relationship can be understood by noting that SEP intensities have a positive correlation with CME speeds, i.e., lower-speed CMEs produce smaller SEP intensities. Therefore, the small SEP events do not fit to the anticorrelation relationship found between spectral index and CME speed for large SEP events. We also studied the correlations between CME initial speed and CME space speed and initial acceleration, and found moderate correlation coefficients of 0.42 and 0.45, respectively. The correlation coefficients are significantly lower than those (0.81 and 0.71, respectively) found by Gopalswamy et al. (2016). A much smaller sample size makes it difficult to assess reliably the significance of the difference but considering that the range of CME acceleration and speed values is narrower for small SEP events than for large SEP events, smaller correlation coefficients for small SEP events are an expected result.

Gopalswamy et al. (2017) studied large SEP events using type II radio bursts and found that the starting frequency of type II bursts organizes SEP events: their results showed that FE-associated SEP events have the lowest average starting frequency ( ~22 MHz), the regular SEP events have an intermediate average (~81 MHz) and the GLEs have the highest average (~107 MHz). They suggested that this result gives further support for the importance of CME kinematics in organization of SEP events. As shown in Fig. 3, we found in our study that the average (median) starting frequency of type II bursts with a clear F-H structure (11 events) for small SEP events is about 76 MHz (80 MHz). The requirement of an existing F-H structure ensures that we can identify reliably the emission lane corresponding to the fundamental plasma frequency. For type II bursts with only a single emission lane we cannot be certain if the observed emission is at fundamental or harmonic frequency, which will increase the scatter of data points. Clearly the starting frequency in small SEP events is close to that of the regular large SEP events. Therefore, small SEP events seem to match the hierarchical structure found by Gopalswamy et al. (2016, 2017).

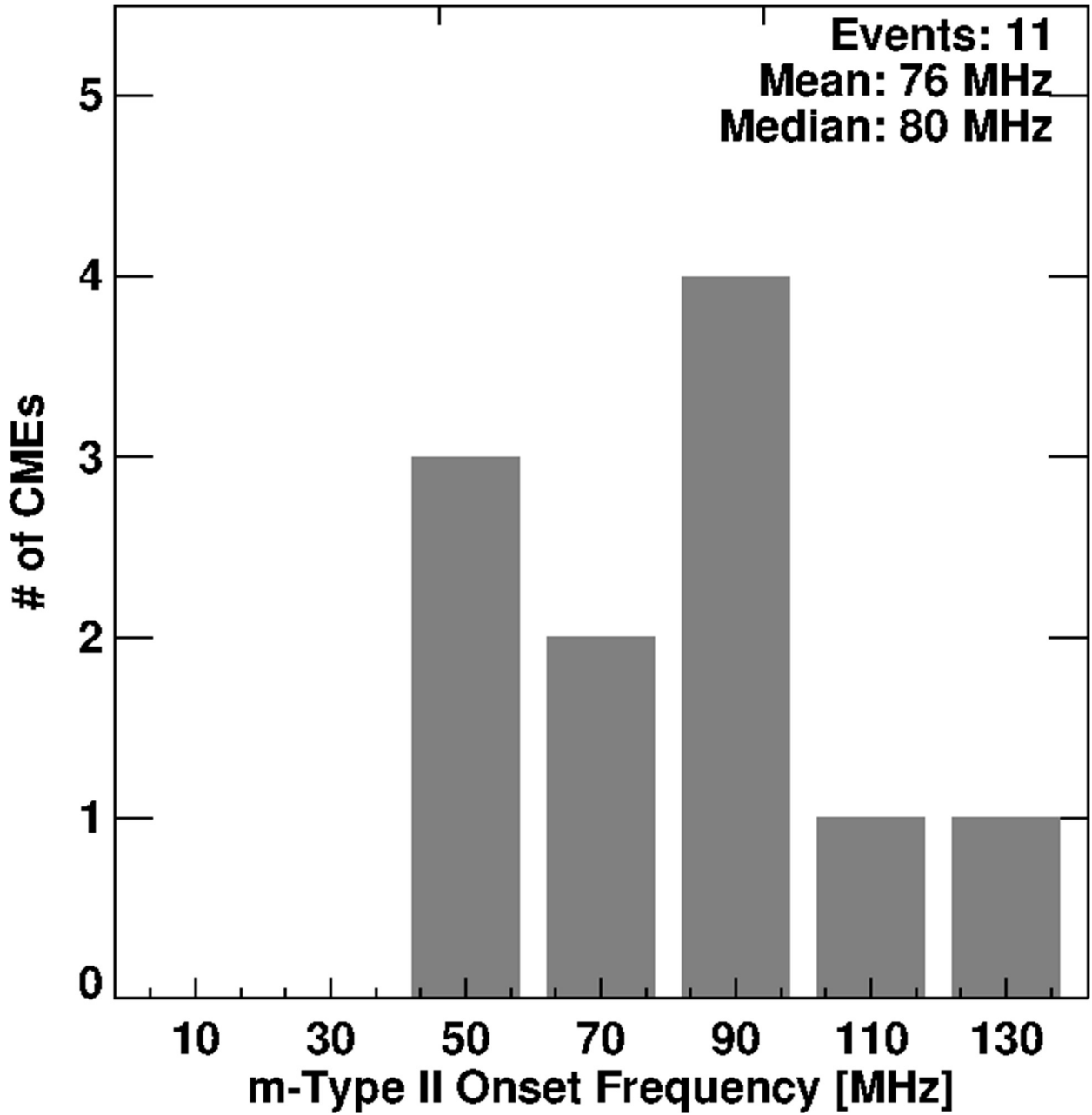

**Figure 3** The distribution of the starting frequency of metric type II bursts with a clear F-H structure associated with small SEP events.

Based on the hierarchical relationship by Gopalswamy et al. (2017) discussed in the previous section, we have studied if the starting frequency of type II bursts organizes the longitudinal extent of SEP events. In our analysis, we have excluded the 23 February 2011 type II burst, because it had an extremely high starting frequency of 425 MHz. The details of this event are discussed by Cho et al. (2013).

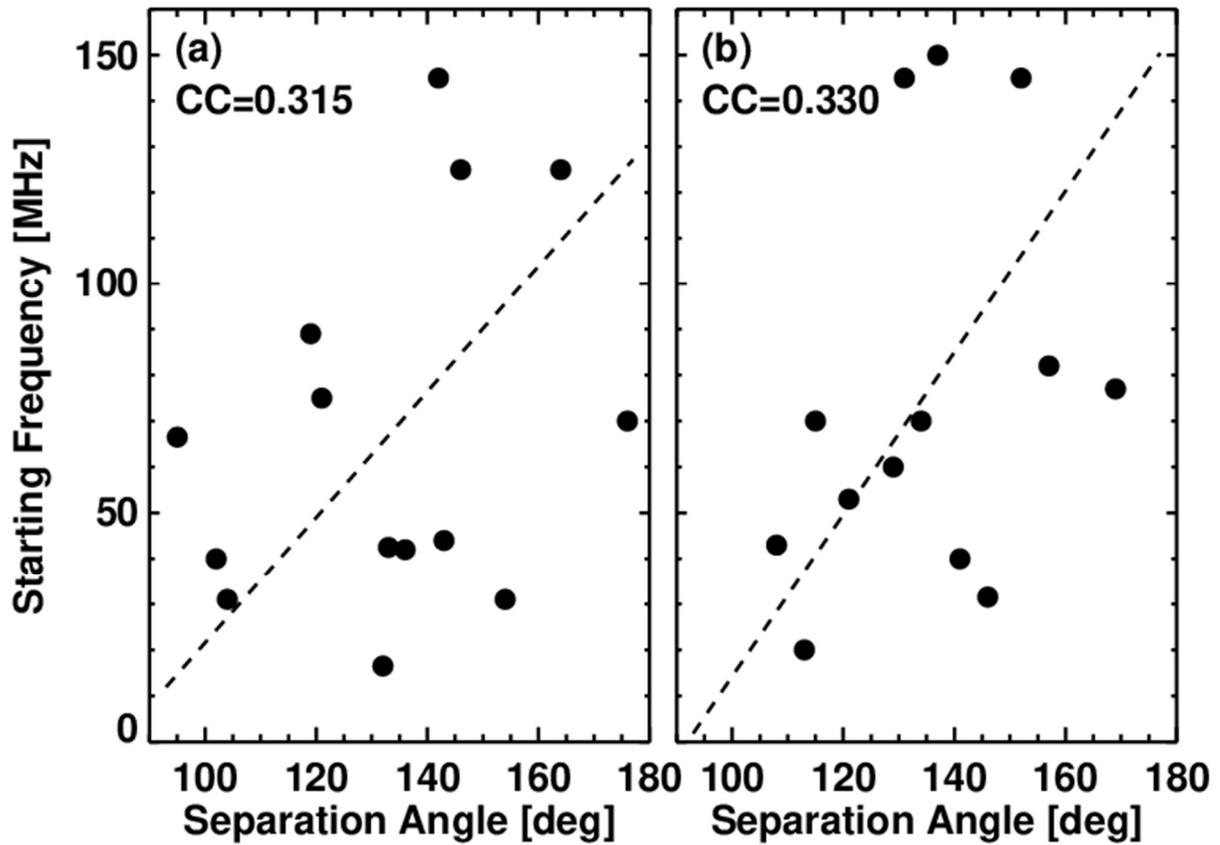

**Figure 4** Starting frequency vs. separation angle for 1-spacecraft (a) and 2-spacecraft (b) > 25 MeV SEP events. All metric type II burst were required to have a F-H structure. Dashed line shows the fitted linear trend.

The results of our analysis of are shown in Figs 4 and 5. Figure 4a shows the measured starting frequency of the type II burst as a function of the separation angle for 1-spacecraft events associated with a metric type II bursts showing a clear F-H structure. Linear fitting to data points in Fig. 4a indicates a positive trend but the correlation coefficient is low, 0.315. Figure 4b shows a similar plot for 2-spacecraft events. Again, linear fitting indicates a positive trend, but the correlation coefficient is 0.330, about the same as in Fig. 4a.

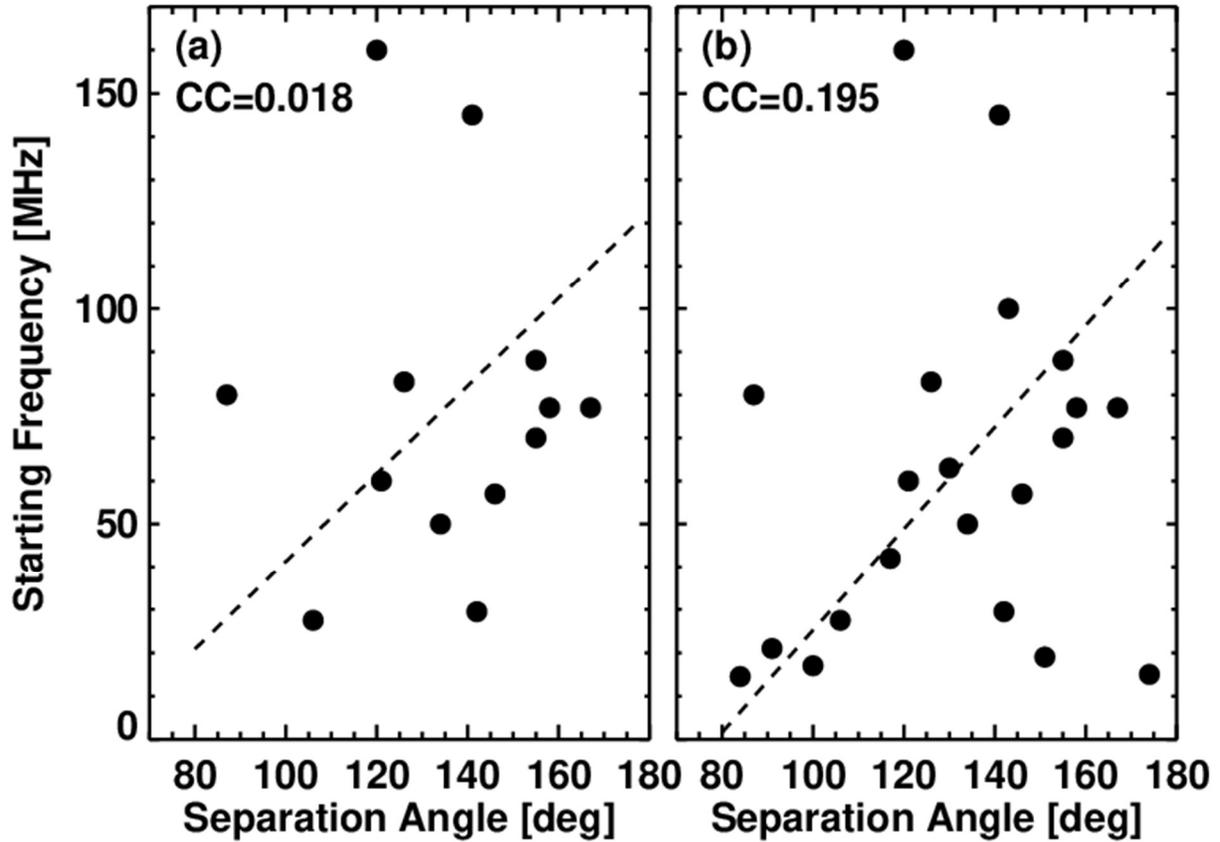

**Figure 5** Starting frequency vs. separation angle for 3-spacecraft >25 MeV SEP events. All metric type II bursts were required to have a F-H structure in panel (a) but in panel (b) we have added events with single-lane type II bursts assuming that the radio emission was at the harmonic frequency. Dashed line shows the fitted linear trend.

In Fig. 5a, we have plotted the starting frequencies for 3-spacecraft events as a function of the separation angle, which in this case is only a lower limit, because all 3 spacecraft detected the same SEP event and the extent is therefore unmeasurable. The correlation coefficient is only 0.018, because there appears to be two outliers with higher starting frequency. If there is no confusion between fundamental and harmonic frequencies, the outliers are most likely caused by the complex high-density structures in the low corona near the solar surface that increase the starting frequency of type II emission. Depending on the detailed structure of the low corona, the shock front could cross coronal loops with higher than average density as was the case for the excluded 23 February 2011 type II bursts with an extremely high starting frequency (Cho et al., 2013).

In Fig. 5b, we have tried to improve the statistics by assuming that all single-lane emission events we excluded in Fig. 5a were observed at the harmonic frequency, i.e., we have divided the measured frequency by two. Scatter of data points remains large and the correlation coefficient is low, 0.195. Linear fitting shows a positive trend for both data sets in Figs 5a,b. If we remove the two possible outliers with starting frequencies around 150 MHz, the correlation coefficients of Figs. 5a,b improve slightly to 0.235 and 0.313, respectively. It is also possible that the two single-lane events in the lower righthand corner of Fig. 5b are events were the type II emission was at the fundamental emission and therefore the plotted frequency should be multiplied by two, which moves the data points closer to the fitted line. In general, we do not find a clear strong linear relationship of the starting frequency of metric type II radio bursts with the separation angle, although linear fitting indicates a similar positive trend for all three data sets. The linear trend is best seen in the Fig. 5b, where most points are close to the dashed line. Most likely the scatter of data points is large due to complex density structures (loops) of the low solar corona, which cannot be captured by our simple analysis method. The high starting frequency of the type II burst indicates that the shock forms low in the corona and therefore the CME must have high acceleration and most likely it reaches a high speed. We know that type II radio bursts are well associated with large SEP events, especially DH type II radio bursts, which are known to be produced by fast and wide CMEs (Gopalswamy et al., 2001). Therefore, we expect the high starting frequency of a metric type II burst to correspond to a longitudinally widespread SEP event, i.e., the positive trend seen in Fig. 5b.

## Conclusions

We found that the average power-law index of fluence spectra in small SEP events is similar to that in large SEP events. The average onset frequency of type II radio bursts is also similar to that of large SEP events. Therefore, the small SEP events show no deviations from the hierarchy of spectral index or the hierarchy of the starting frequency of type II radio bursts previously established for the large SEP events. This is an important result indicating that the physics of particle acceleration is the same in small and large SEP events.

Linear fitting indicates a positive trend between the onset frequency of metric type II bursts and the estimated separation angle for all sets of SEP events, but the scatter of data points is large, and the correlations are low. Because the number of spacecraft locations where SEP measurements are made is very limited, the longitudinal extent of the SEP event at 1 AU is currently difficult to estimate. The starting frequency of type II radio bursts is not affected only by the CME kinematics, but also by high-density structures in the low corona, which increases the scatter of data points.

## Acknowledgements

We thank the science teams of the SDO, SOHO, and STEREO missions and ground-based radio observatories for making their SEP, CME and solar radio burst observations available. The work of PM and NT was partially supported by NASA grant NNX15AB77G. HX was supported by NASA LWS grant NNX15AB70G and NASA HGI grant NNX17AC47G. The work of NG was supported in part by the NASA LWS program. SOHO is a project of international collaboration between ESA and NASA.## References

Baker, D.N., Mason, G.M., Figueroa, O., Colon, G., Watzin, J.G., Aleman, R.M.: 1993, IEEE T. Geosci. Remote (ISSN 0196-2892) 31, 531, DOI:10.1109/36.225519.

Bougeret, J.-L., Kaiser, M.L., Kellogg, P.J., et al.: 1995, Space Sci. Rev. 71, 231, DOI:10.1007/BF00751331.

Brueckner, G.E., Howard, R.A., Koomen, M.J., et al.: 1995, Solar Phys. 162, 357, DOI:10.1007/BF00733434.

Cane, H.V.: 1996, AIP Conf. Proc. 374, 124, DOI:10.1063/1.50948.

Cane, H.V., and Stone, R.G.: 1984, Astrophys. J. 282, 339, DOI:10.1086/162207.

Cho, K.-S., Gopalswamy, N., Kwon, R.-Y., Kim, R.-S., Yashiro, S.: 2013, Astrophys. J. 765, 148, DOI:10.1088/0004-637X/765/2/148.

Cliver, E.W., Kahler, S.W., Reames, D.V.: 2004, Astrophys. J. 605, 902, DOI:10.1086/382651.

Cliver, E.W., Thompson, B.J., Lawrence, G.R., et al.: 2005, in B. Sripathi Acharya, S. Gupta, P. Jagadeesan, A. Jain, S. Karthikeyan, S. Morris, and S. Tonwar. (eds), Proc. 29th Int. Cosmic Ray Conf. Vol. 1, Tata Institute of Fundamental Research, Mumbai, India, p. 121.
21